\documentclass[review]{elsarticle}
\usepackage{mathpazo}

\usepackage[T1]{fontenc}
\usepackage[latin9]{inputenc}
\usepackage{fancyhdr}
\usepackage{float}
\usepackage{amstext}
\usepackage{graphicx}
\usepackage{esint}
\usepackage{url}
\usepackage[unicode=true,
 bookmarks=true,bookmarksnumbered=false,bookmarksopen=false,
 breaklinks=false,pdfborder={0 0 1},backref=false,colorlinks=false]
 {hyperref}
\usepackage{breakurl}

\makeatletter

\providecommand{\tabularnewline}{\\}

\@ifundefined{date}{}{\date{}}

\usepackage{lipsum}

\usepackage{microtype}

\usepackage{multicol}
\usepackage[hang, small,labelfont=bf,up,textfont=it,up]{caption}
\usepackage{booktabs}
\usepackage{float}

\usepackage{lettrine}
\usepackage{paralist}

\usepackage{abstract}

\usepackage{titlesec}
\renewcommand{\thesection}{\Roman{section}}
\titleformat{\section}[block]{\large\scshape\centering}{\thesection.}{1em}{} 

\makeatother

\defcitealias{Commission2014}{European Commission}
\defcitealias{Directorate2014}{NPD}

\begin{document}
\thispagestyle{fancy}

\title{Forecasting future oil production in Norway and the UK: a general improved methodology}

\author[eth-mtec]{L. Fi\'evet\corref{cor1}}
\ead{lfievet@ethz.ch}

\author[eth-mtec]{Z. Forr\'o}
\ead{zforro@ethz.ch}

\author[eth-mtec]{P. Cauwels}
\ead{pcauwels@ethz.ch}

\author[eth-mtec]{D. Sornette}
\ead{dsornette@ethz.ch}

\cortext[cor1]{Corresponding author. Tel.: +41 44 632 83 79.}

\address[eth-mtec]{Chair of Entrepreneurial Risks, ETH Z\"urich, Scheuchzerstrasse 7 (SEC F), CH-8092 Z\"urich, Switzerland}

\maketitle
\begin{abstract}
\noindent We present a new Monte-Carlo methodology to forecast
the crude oil production of Norway and the U.K. based on a two-step process,
(i) the nonlinear extrapolation of the current/past
performances of individual oil fields and (ii) a stochastic model
of the frequency of future oil field discoveries. Compared
with the standard methodology that tends to underestimate remaining oil reserves,
our method gives a better description of future oil production, as validated by
our back-tests starting in 2008. Specifically, we predict remaining reserves
extractable until 2030 to be $188 \pm 10$ million barrels for Norway and $98 \pm10$ million
barrels for the UK, which are respectively $45\%$ and $66\%$ above the
predictions using the standard methodology.
\end{abstract}

\section{Introduction}

\lettrine[nindent=0em,lines=3] Forecasting future oil production
has been a topic of active interest since the beginning of the past
century because of oil central role in our economy. Its importance
ranges from energy production, manufacturing to the pharmaceutical
industry. As petroleum is a non-renewable and finite resource, it
is primordial to be able to forecast future oil production. The fear
of a global oil peak, beyond which production will inevitably decline,
has been growing due to stagnating supplies and high oil prices since
the crisis in 2008/2009 \citep{Murray2013}. As any industrialized
country, Europe is strongly dependent on oil supply to maintain its
economic power. In the nowadays difficult geopolitical environment,
it is important to know how much of the oil needed in Europe will
come from reliable sources. In the past, a big share has been coming
from Norway and the U.K., two of Europe's biggest exporters. However,
the U.K. already became a net importer in 2005 and Norway\textquoteright{}s
production has been declining rapidly as well \citep{Hook2008}.\\
 The methodology behind forecasting future oil production has not
evolved much since M. King Hubbert, who in 1956 famously predicted
that the U.S. oil production would peak around 1965-1970 \citep{Hubbert1956}.
That prediction has proven itself to be correct. His main argument
was based on the finiteness of oil reserves and to what amounts
as the use of the logistic differential equation for the total
quantity $P(t)$ of oil extracted up to time $t$
\begin{equation}
\frac{dP}{dt}=rP\left(1-\frac{P}{K}\right)~.
\label{logistic}
\end{equation}
The logistic differential equation
is characterized by an initial exponential growth, which then decreases
to zero as the total oil extracted reaches saturation (no more oil
is to be found). The parameter $r$ is commonly referred to as the
growth rate, and $K$ as the carrying capacity (total quantity of
oil that can be ultimately extracted). If $P(t)$ is the amount of oil extracted up to
time $t$, then $f(t) := \frac{dP}{dt}$ is the oil production rate, the quantity
that M. King Hubbert predicted with surprising accuracy to peak. From
a methodological point of view, the Hubbert model has enjoyed a longstanding
popularity in modeling future oil production given its simplicity.
Various extensions have been studied by \citet{Brandt2007}
to account for multi-cycled or asymmetric production curves.
The existing forecasts of future oil production use some form of the
Hubbert model \citep{Brecha2012,Laherrere2002,Lynch2002}
or some economical model applied to aggregate production \citep{Greiner2011},
but none goes into the details of studying the underlying dynamics.
The main reason for the lack of details is certainly the lack of available
data. 

In this article, a new methodology is introduced to forecast future
oil production. Instead of taking the aggregate oil production profile
and fitting it with the Hubbert curve or its variants (such as the
multi-cyclic Hubbert curve), the production profile of each individual
oil field is used. By extending their production into the future and
extrapolating the future rate of discovery of new fields, the future
oil production is forecasted by means of a Monte Carlo simulation.
To demonstrate the generality of the methodology presented here,
it is applied to two major oil producing countries with publicly available
data: Norway \citepalias{Directorate2014} and the U.K. \citep{GOVUK2014}.

\section{Methodology}

The idea behind the methodology is to model the future aggregate oil
production of a country by studying the production dynamics of its
individual constituents, the oil fields. The main benefit of this
approach, compared to working directly with aggregate production data,
is the possibility to forecast non-trivial oil production profiles
arising from the combination of all the individual field
dynamics. This ability reflects directly the fact that the total
production is the sum of the contributions of each individual
oil field in production. Thus modelling at the level of each field
reflects more closely the reality and is likely to be over-performing
and more reliable, as we show below.
In order to implement our approach, one must be able 1) to \textbf{extend
the oil production of each individual field} into the future and 2)
to \textbf{extrapolate the rate of discoveries} of new oil fields.

\subsection{Extending the oil production of individual fields}

The first step to predicting the future oil production of a country
is to extrapolate the future production of existing fields and to
estimate the error on this extrapolation. The data of the fields developed
in the past shows a repeating asymmetric pattern. A good example is
the Oseberg field shown in figure \ref{regular}, with a quick ramp
up once the field is being developed, and then a peak or plateau before
the oil field production starts decaying. The decay can take many different shapes
and is governed by a variety of geological and economical factors.
The goal of the fitting procedure is to capture as much the impact of
these different factors as possible.

\subsubsection{Regular, irregular and new fields\label{sub:Regular,-irregular-and}}

\begin{figure}[H]
\includegraphics[width=1\columnwidth]{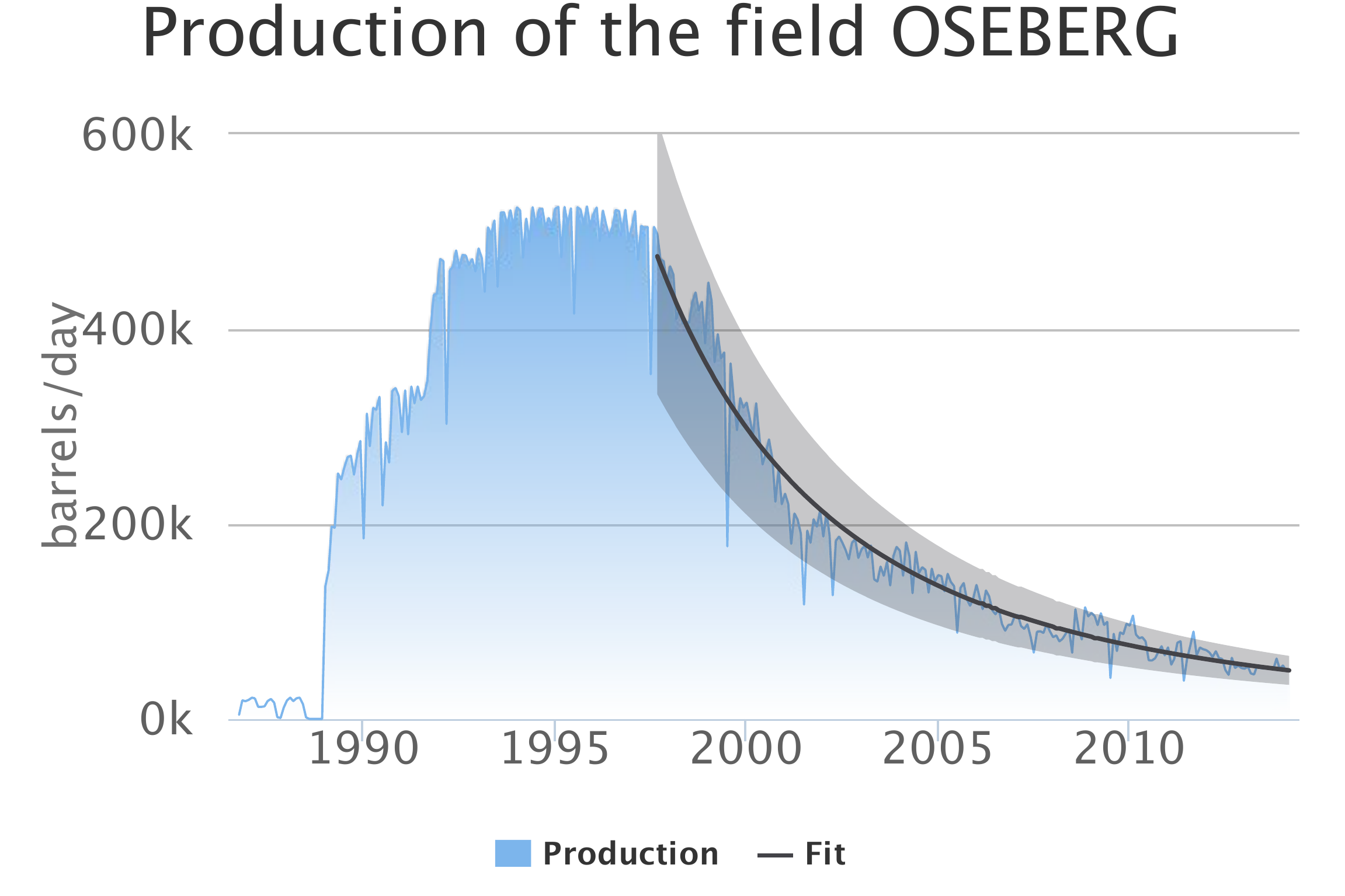}
\caption{Example of the time evolution of the oil production per day of
a regular field, parameterised beyond the peak in the decay
regime by the stretched exponential 
function (\ref{stretched_exponential}) with $\beta=0.66\pm0.01$ and $\tau=-55\pm1$ months,
shown with the black line.  The one standard deviation given by expression
(\ref{eq:extrapolation}) is represented by the grey band.}
\label{regular} 
\end{figure}

\begin{figure}[H]
\includegraphics[width=1\columnwidth]{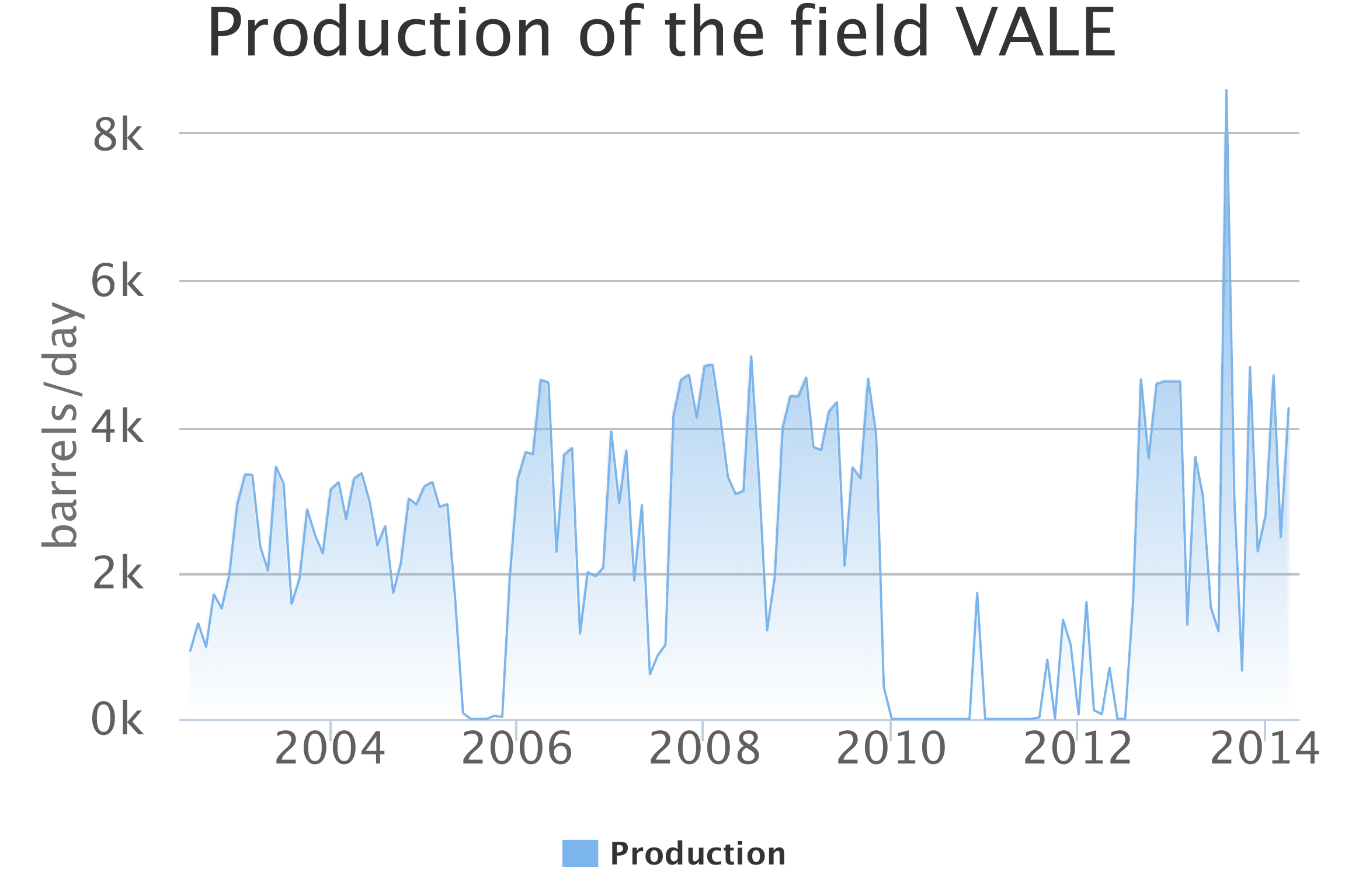}
\caption{Example of a irregular field.}
\label{irregular} 
\end{figure}

\begin{figure}[H]
\includegraphics[width=1\columnwidth]{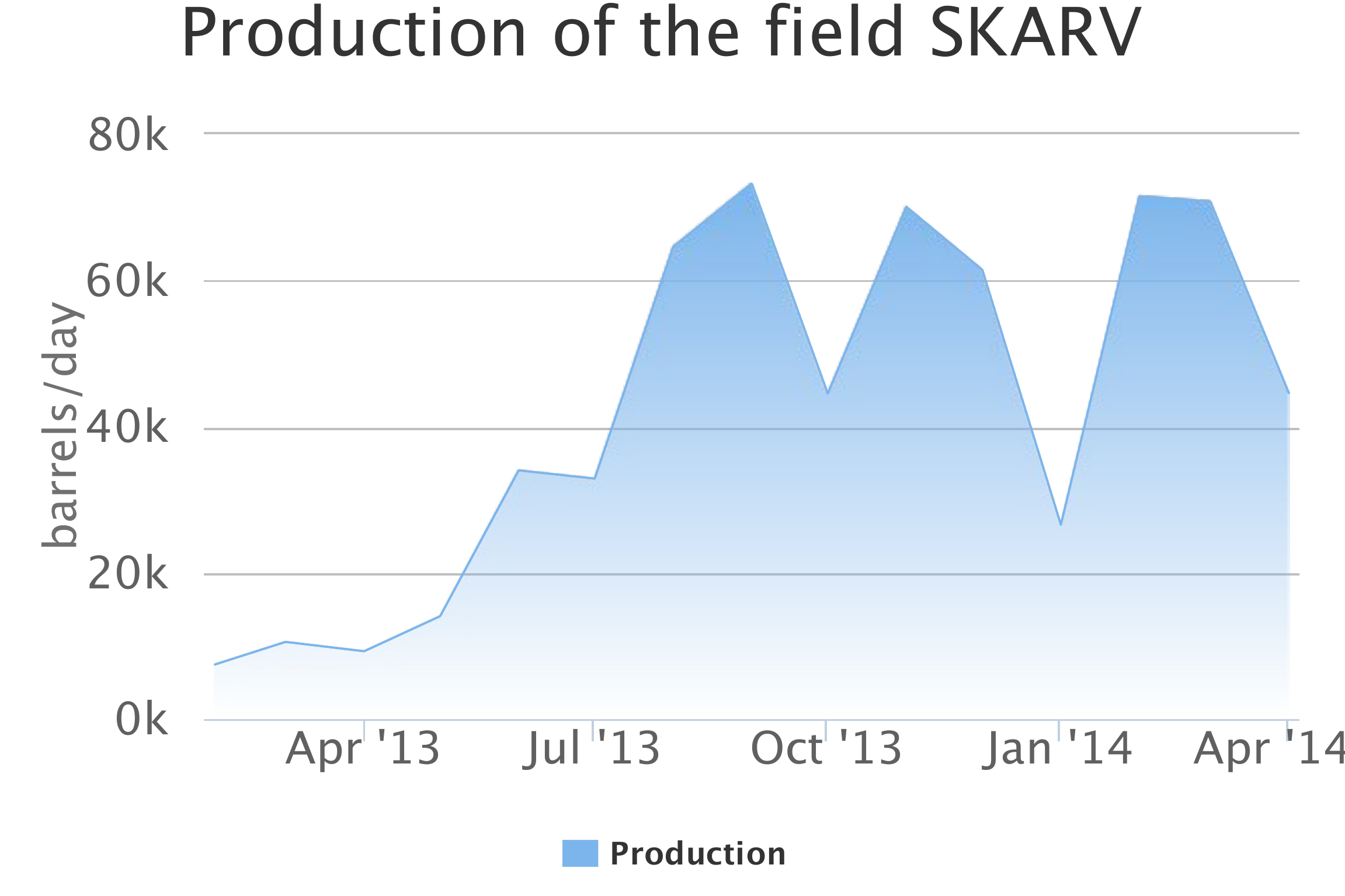}
\caption{Example of a new field.}
\label{new} 
\end{figure}

To be able to forecast the oil production of each individual field,
regularity had to be found in the production's dynamics. Modeling
the whole production profile from the beginning of extraction seems
elusive due to the variety of the forms it can take. Fortunately,
modeling the decay process is sufficient in order to extrapolate future
oil production. A preliminary classification is necessary to achieve
that goal. Figures \ref{regular}, \ref{irregular} and \ref{new}
show that, independent of the country, oil fields can be classified
into three main categories:
\begin{itemize}
\item \textbf{Regular fields} - Their decays show some regularity (see figure \ref{regular} ); 
\item \textbf{Irregular fields} - The ones that do not decay in a regular
fashion (see figure \ref{irregular});
\item \textbf{New fields} - The ones that do not decay yet. As such, there
is no easy way to forecast their future oil production based on past
data (see figure \ref{new}).
\end{itemize}
\noindent All the fields have been fitted using an automated algorithm,
but the results have been subsequently checked visually to sort out
the irregular fields which could not be fitted. As of January 2014,
regular fields make up 85\% and 87\% of the number of fields and 94\%
and 71\% of the total produced oil volume in Norway and the U.K. respectively.
As such, being able to model them is crucial. To capture as many different
decay dynamics as possible, the decay part of the oil production rate
$f(t) := \frac{dP}{dt}$ has been fitted by the stretched exponential  
\begin{equation}
f(t)=f_{0}~e^{-\left(\frac{t}{\tau}\right)^{\beta}}~.
\label{stretched_exponential}
\end{equation}
The stretched exponential function
has many advantages as it generalises the power law and can therefore
capture a broad variety of distributions as shown by \citet{Laherrere1998}. Moreover,
\citet{Malevergne2005} showed that the power law
function can be obtained as an asymptotic case of the stretched
exponential family, allowing for asymptotically nested statistical tests.  As can be seen
in figure \ref{regular}, the stretched exponential (equation \ref{stretched_exponential})
is a good functional form to fit the decay process of regular fields.

For the minority of irregular fields, we assume 
no difference in the decay of production between giants and dwarfs, which
has been modelled as follows: the decay time scale $\tau$ has been picked to be the average $\tau$ over
the regular fields. Then $\beta$ has been fixed so that the sum of
the field production over its lifetime be equal
to the official ultimate recovery estimates, when such an estimate
is available.

The minority of new fields, which did not yet enter their decay phase,
cannot be extrapolated and will therefore be treated as new discoveries.
The technical details of how to treat them as new discoveries are
discussed in section \ref{sub:Future-production-from}.

\subsubsection{Back-testing \& Error}

\begin{figure}[H]
\includegraphics[width=1\columnwidth]{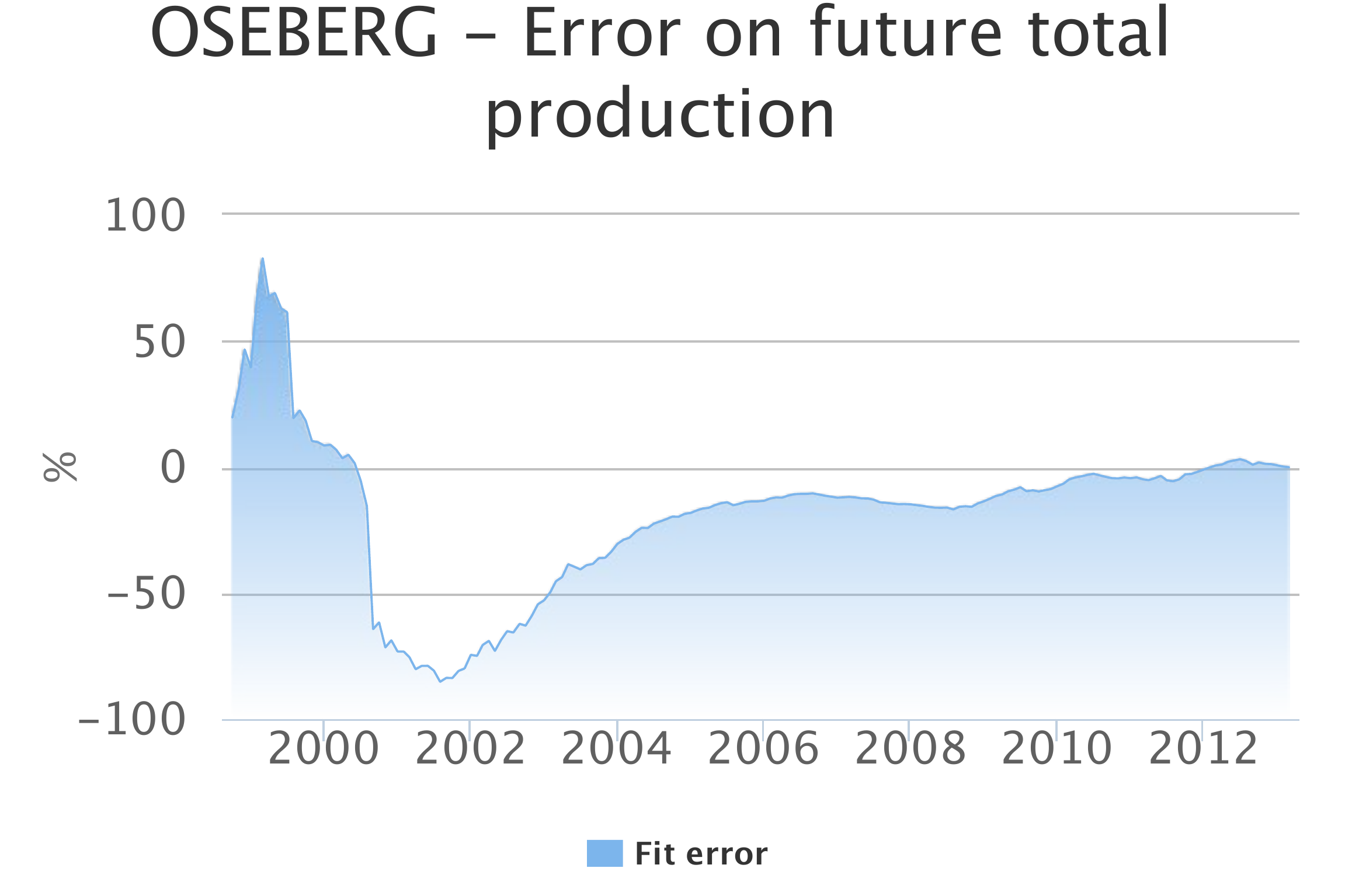}
\caption{Oesberg field - Relative error defined by expression (\ref{ryjryukoik})
of the predicted total production from time $t$ indicated in the abscissa until 2014.
One can observe that the predicted future total production is over-estimated by as much
as 70\% in 1999, then under-estimated by the same amount in 2002, while the forecast
errors remain smaller than 20\% since 2004.}
\label{oesberg-error}
\end{figure}

To determine how well the extrapolation based on the stretched exponential
predicts the future production, a complete back-testing has been performed
on each field. A single back-test is made as follows:
\begin{itemize}
\item The production data $\{p_{0},\,\ldots,\, p_{N}\}$ of the field is
truncated at a certain date in the past $T \in\{0,\,\ldots,\, N\}$, where
$T$ is the time counted in months since the production start of the field.
\item The extrapolation of the oil production rate
$f(t) := \frac{dP}{dt}$ is made based on the truncated data $\{p_{0},\,\ldots,\, p_{T}\}$.
\item The future production predicted by the extrapolation function $f(t)$ can be
compared to the actual production from the date $T$ in the past
up to the present $T_{f}=N$. The extrapolated total production can be
computed as 
\begin{equation}
P_{e}(T)=\int_{T}^{T_{f}}f(t)dt
\end{equation}
and the relative error is given by
\begin{equation}
e(T)=\frac{P_{e}(T)-\sum_{i=T}^{T_{f}}p_{i}}{P_{e}(T)},
\label{ryjryukoik}
\end{equation}
where both $P_{e}(T)$ and $e(T)$ are functions of the truncation time $T$.
\end{itemize}
Computing this back-test, for every month in the past since the field production
started decaying, yields a plot showing the evolution of the relative
error over time defined by expression (\ref{oesberg-error}). By construction, the relative
error will tend to zero as the truncation time $T$ approaches the
present. Nonetheless, it is a useful indicator for the stability of
the extrapolation. As can been seen for the Oseberg field in figure \ref{oesberg-error}, the relative
error on future production remained fairly stable during the past
decade.

From the complete back-test, we compute the average relative error
\begin{equation}
\bar{e}=\frac{1}{N}\sum_{i=0}^{N}e(i)
\end{equation}
of the extrapolation made on the future production. Assuming that the
relative errors are normally distributed around the average relative
error, the standard deviation on the average relative error is given
by 
\begin{equation}
\sigma_{e}=\sqrt{\frac{1}{N}\sum_{i=0}^{N}\left(e(i)-\bar{e}\right)^{2}}.\label{eq:field-risk}
\end{equation}
As the average relative error is often fairly constant, the extrapolation
was corrected by the average relative error, that is, if the extrapolation
consistently over-estimated the production by 10\% during the back-test,
the extrapolation was reduced by 10\%. This results in an extrapolated
production $p(t)$, including a 1$\sigma$ confidence interval, given by
\begin{equation}
p(t)=(1-\bar{e})\, f(t)\pm\sigma_{e}.
\label{eq:extrapolation}
\end{equation}
An example of such an extrapolation including a one standard deviation
range is shown in figure \ref{regular} for the Oseberg field.

\subsubsection{Aggregate error}

Once the individual fields have been extrapolated using
formula \ref{eq:extrapolation}, we compute the extrapolation of the oil production
for the whole country. While it is straightforward to sum the extrapolations
of the individual fields to obtain the expected production, some care
has to be taken with respect to the confidence interval of the production
at the country level.

As shown later in section \ref{sec:Results}, the same extrapolation
including a complete monthly back-test of total future production
has been performed at the country level and the resulting relative
error is much smaller than the average error observed on the individual
fields. To account for this observation, the assumption made is that
the relative error between individual fields is uncorrelated. 
While one could imagine that some inter-dependence could result from
a coordinated response of supply to a sharp increase/decrease of demand,  we have not 
observed this to be the case at a significant level. Therefore,
the fields can be considered as a portfolio of assets with a return
given by their extrapolation $p(t)$ (eq. \ref{eq:extrapolation})
and a risk given by $\sigma_{field}$ (eq. \ref{eq:field-risk}).
This means that the average standard deviation per field
at the country level from the extrapolated production can be computed as 
\begin{equation}
\sigma_{country}^{2}=\frac{1}{\#fields}\sum_{field\in fields}\sigma_{field}^{2}.
\end{equation}
Intuitively, this models well the fact that the uncorrelated errors
among fields will mostly cancel out.

\subsection{Discovery rate of new fields\label{sub:Discovery-rate-of}}

Knowing the future production rate of existing fields is not enough
as new fields will be discovered in the future. The model describing
the discovery rate of new fields should satisfy two fundamental observations.
\begin{enumerate}
\item The rate of new discoveries should tend to zero as time goes to infinity.
This is a consequence of the finiteness of the number of oil fields. 
\item The rate of new discoveries should depend on the size of the oil fields.
As of today, giant oil fields are discovered much less frequently
than dwarf oil fields.
\end{enumerate}

\subsubsection{Discoveries modeled as logistic growth\label{sub:Discoveries-modeled-as}}

\begin{figure}[H]
\includegraphics[width=1\columnwidth]{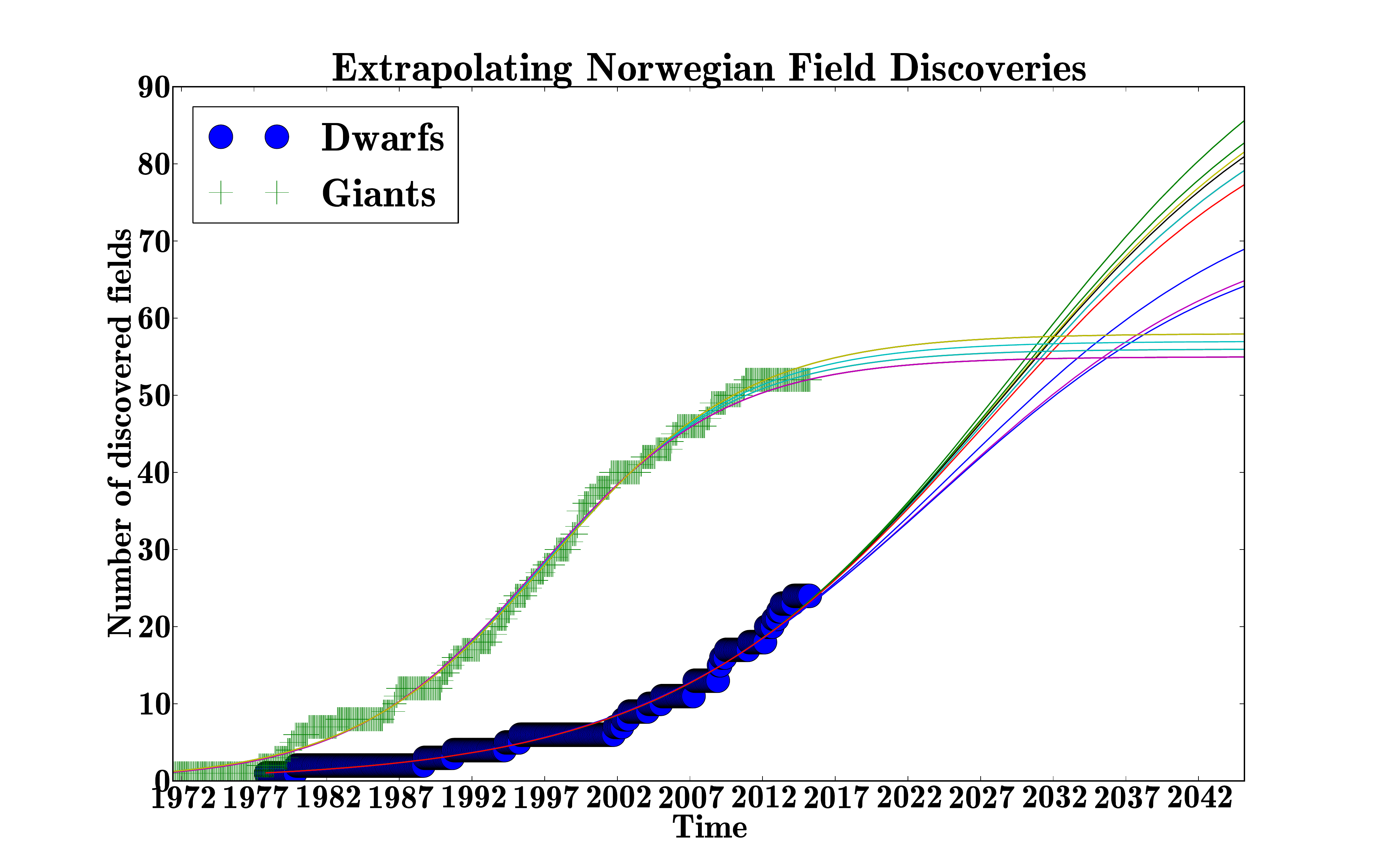}
\caption{Logistic fit of the function solution of expression (\ref{logistic})
to the number of discoveries for Norway. The discovery
rate of new oil fields is dependent on their size as explained in the main text.}
\label{rate_of_discovery} 
\end{figure}

A natural choice for such a model is a non-homogenous Poisson process.
The Poisson process is a process that generates independent events
at a rate $\lambda$. It is nonhomogeneous if the rate is time-dependent,
$\lambda\rightarrow\lambda(t)$. The standard way to measure $\lambda(t)$
is to find a functional form for $N(t)$, the statistical average
of the cumulative number of events (discoveries) up to time $t$.
Then, $\lambda(t)$ is simply a smoothed estimation 
of the observed rate $\frac{dN(t)}{dt}$. Figure \ref{rate_of_discovery}
shows $N(t)$ for Norwegian fields classified according to their size in two classes, 
dwarfs and giant fields. 
The logistic curve is a good fit to the data (integral form of equation
\ref{logistic}). This implies that after an initial increase, the
rate of new discoveries reaches a peak followed by a decrease until
no more oil fields are to be found, consistent with our fundamental
observations. This same approach has already been successfully applied
by \cite{Forro2012} to estimate the number of daily active
users on Zynga.

As the discovery and production dynamics are not independent of the
field size, the fields have been split into two groups: dwarfs and
giants. Unfortunately, the two logistic curves thus obtained are highly
sensitive to the splitting size. This results from the major issue,
when fitting a logistic curve to data, that the carrying capacity
can not be determined if the data does not already exhibit the slowdown 
in growth towards the carrying capacity. However, it is
mentioned in the literature that often dwarf fields have already been
discovered a long time ago, but their production has been postponed
for economical reasons \cite[p. 378]{Lynch2002}. Therefore, it is
expected that the large oil fields have mostly been found and produced,
and that future discoveries will mostly be made up of dwarf fields.
Consequently, the splitting size has been picked as small as possible
in order to maximize the number of giant fields but large enough to
avoid recent discoveries. Our definition is thus:
\begin{itemize}
\item \textbf{Dwarfs}: Fields which produced less than $50\cdot10^{6}$
barrels.
\item \textbf{Giants}: Fields which produced more than $50\cdot10^{6}$
barrels.
\end{itemize}
We note that this definition differs by a factor 10 from the more standard one, for which
oil fields with an ultimate recoverable resource of 0.5 billion barrels (Gb) 
or higher are classified as giants, while oil fields with smaller URR are considered to be dwarfs
\citep{Hook2008}.

The resulting plot shown in figure \ref{rate_of_discovery} pictures
the dynamics: giant oil fields have mostly been found while the discovery
process for dwarf fields is still ongoing. The logistic growth curve
fit to the giant discoveries is well constrained, however the fit
for the dwarfs is poorly constrained. As can be seen in figure \ref{rate_of_discovery},
the carrying capacity $K$ of the logistic growth model is not well
constrained by the available data. A large spectrum of values for
$K$ can lead to an equally good fit of the data. We have not taken
into account the possible effect that
newly discovered dwarf fields could become smaller and smaller 
until the new fields have a size that is too small to be economical to drill.

\begin{figure}[H]
\includegraphics[width=1\columnwidth]{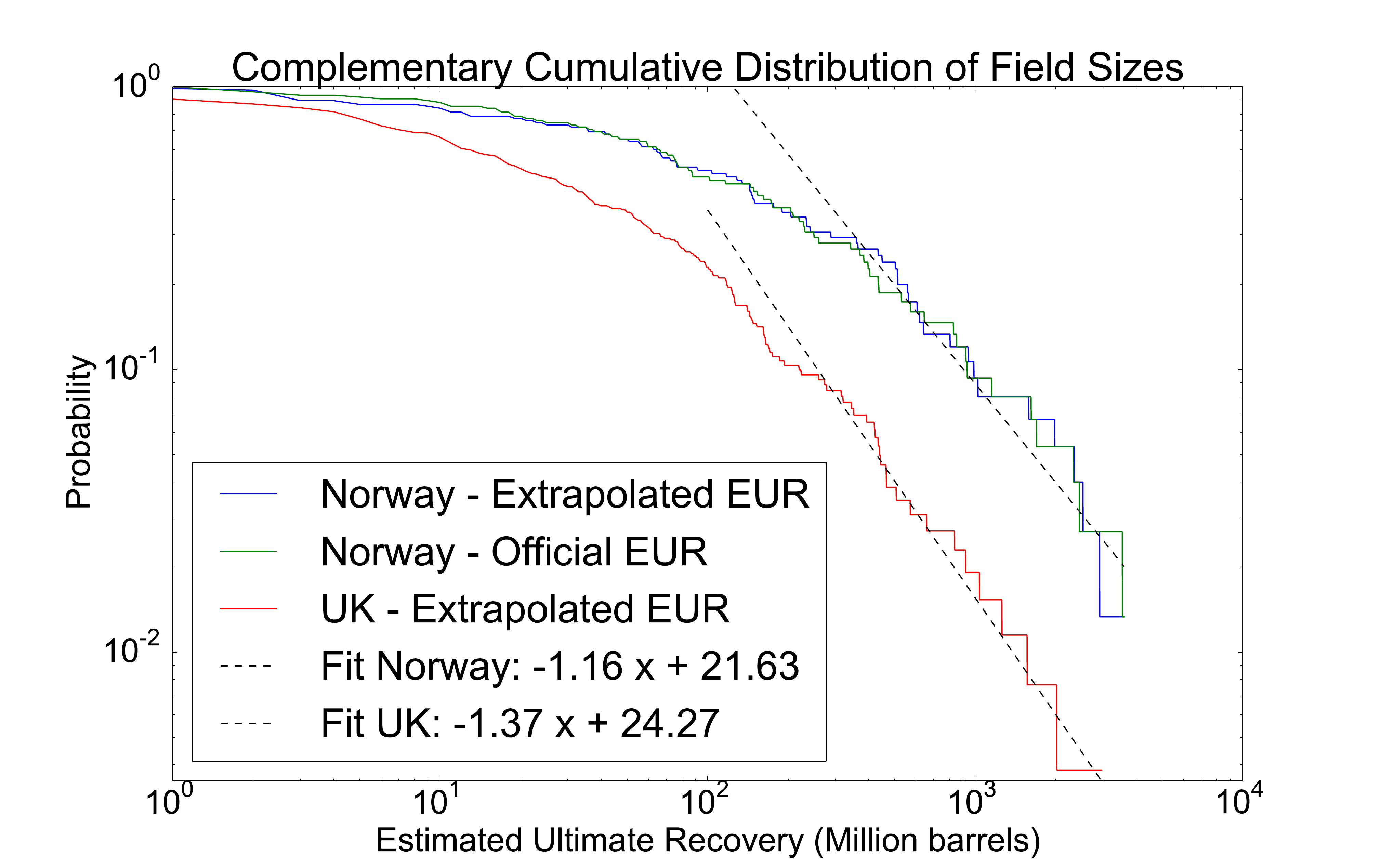}
\caption{Complementary 
cumulative distribution function (CCDF) of known oil field sizes $S$ from Norway and the UK.
The two dashes lines visualise the power law behaviour (\ref{ccdfsizefield}) of the tail
of the distributions. }
\label{ccdffieldsizes}
\end{figure}

There are in fact two competing effects that are likely to compensate each other.
On the one hand, figure \ref{ccdffieldsizes} shows the complementary 
cumulative distribution function (CCDF) of known oil field sizes $S$ from Norway and the UK.
Two salient properties can be observed. First, the tails 
of the distributions are well described by power laws
\begin{equation}
{\rm CCDF}(S) \sim {1 \over S^{1+\alpha}}~,
\label{ccdfsizefield}
\end{equation}
with exponent $\alpha = 1.2 \pm 0.1$ for Norway and $\alpha = 1.4 \pm 0.1$ for the UK.
The fact that the estimations of the exponents $\alpha$ are larger than $1$ implies
that the cumulative oil reserves are asymptotically controlled by the largest fields, and not the small ones
\citep{Sornette2004}. However, the fact that the exponents $\alpha$ are rather close
to $1$ (which is called ``Zipf law'') would make the many small oil field contributing 
significantly in total. This brings us to the second important feature exhibited by figure \ref{ccdffieldsizes},
namely the roll-overs of the CCDFs for small oil fields, likely due to an under-sampling
of the data. Indeed, as for most data sets involving broad distributions of sizes such
as oil fields, the distributions are in general incomplete for the small events due to the non
exhaustive sampling. This incompleteness raises the spectre that our extrapolations
might be grossly underestimating the large potential contributions to the total reserves
of the many small yet undiscovered small oil fields. Assuming that the power law (\ref{ccdfsizefield})
would hold for smaller fields down to size of 1 Million barrels leads to a number of 
such fields 10 to 100 times larger than presently known. 

\begin{figure}[H]
\includegraphics[width=1\columnwidth]{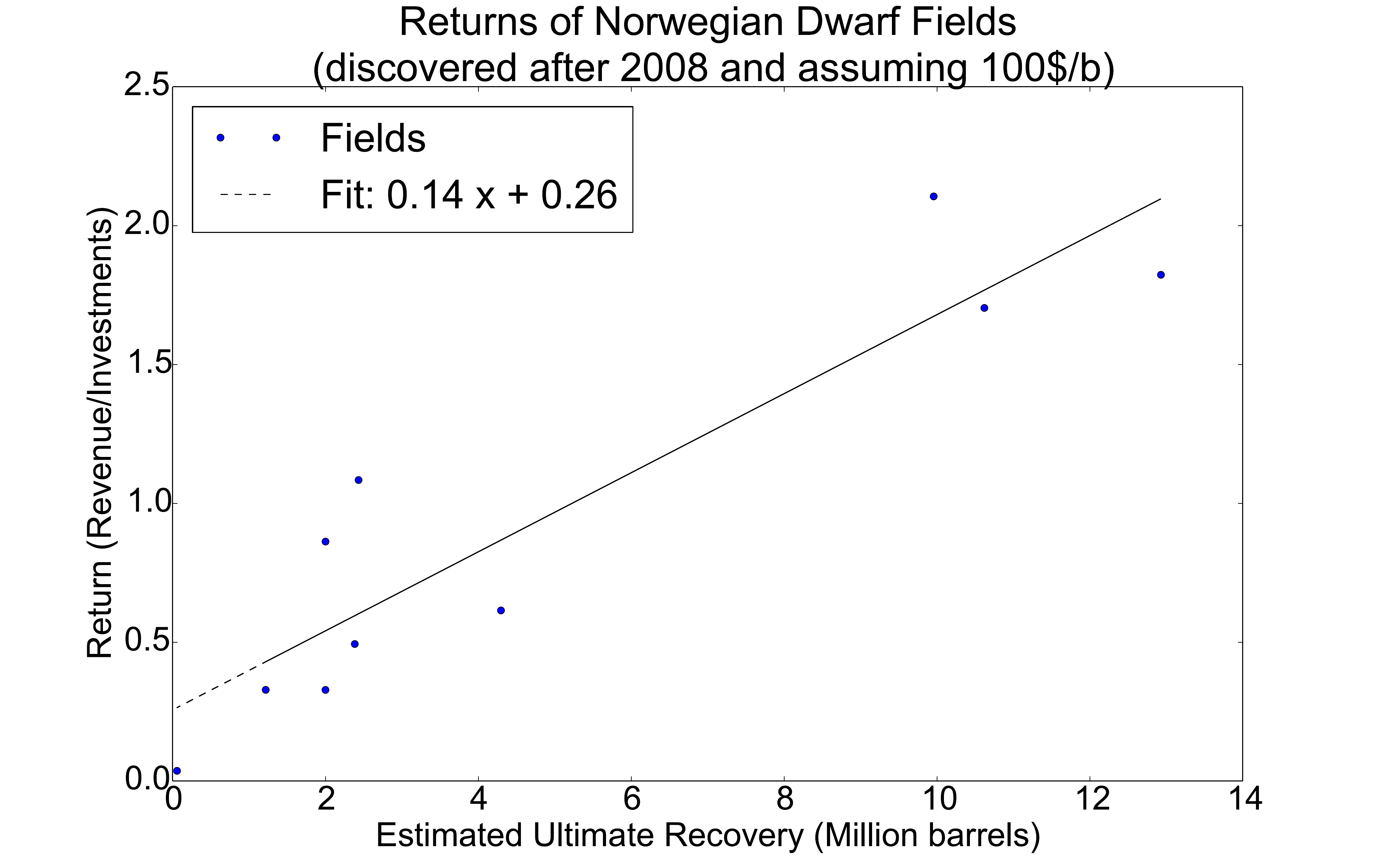}
\caption{Total estimated revenue divided by investment as a function of the estimated 
ultimate recovery for a number of small oil fields. }
\label{returnNorwayfield}
\end{figure}

But there is another
key factor that needs to be considered, namely the fact that small oil fields
are not economically viable for exploitation, which leads to an effective truncation
in the distribution (\ref{ccdfsizefield}) relevant for the estimation of recoverable oil.
Taking the data from the \citetalias{Directorate2014}
providing the yearly investments broken by fields in Norway, let us consider the illustrative
case of the field GAUPE. The investment spent to 
develop its exploitation was  \$380M, while its estimated size is 
$\simeq 1.2Mb$, which corresponds approximately to a total
market value of $\$120M$ at \$100 per barrel. It is thus not a surprise that
investment to exploit this field was interrupted in 2013. 
Figure \ref{returnNorwayfield} shows the total estimated revenue divided
by investment as a function of the estimated ultimate recovery for a number
of small oil fields. It can be inferred that oil fields of sizes smaller than about 10 Mb
are not economically viable as long as the market oil price does not 
grow much higher than \$100 per barrel. This implies that, notwithstanding
the large number of unknown small oil fields, economic considerations
oblige us to neglect the small oil fields, therefore providing
a justification of our procedure. In fact, economic constraints may lead to 
cap the carrying capacity at a value smaller than 
the one shown in figure  \ref{rate_of_discovery}. 

To address these issues from a more solid angle, the method
described in the next section \ref{sub:Likelihood-function-for} has been used
to compute the probability of different carrying capacities.

\subsubsection{Likelihood function for the number of discoveries\label{sub:Likelihood-function-for}}

To overcome the poor constraint on the carrying capacity
$K$ obtained from the fitting procedure for dwarf fields, a method
already used by \cite{Smith1980} has been implemented. This
method makes the following two postulates:
\begin{enumerate}
\item ``The discovery of reservoirs in a petroleum play can be modeled
statistically as sampling without replacement from the underlying
population of reservoirs.''
\item ``The discovery of a particular reservoir from among the existing
population is random, with a probability of discovery being dependent 
on (proportional to) reservoir size.''
\end{enumerate}
The fields are split into $J$ size bins denoted $S_{1},\,\ldots,\, S_{J}$
occurring with frequency $n_{1},\,\ldots,\, n_{J}$. Each discovery
is considered as a step $i$ at which a field of size $I(i)\in\{S_{1},\,\ldots,\, S_{J}\}$
is found and $m_{ij}$ denotes the number of fields of size $j\in\{1,\,\ldots,\, J\}$
discovered before the $i^{th}$ step. Then, the probability that the
discovery at step $i$ is of size $j$ can be expressed as
\begin{equation}
P\left(I(i)=S_{j}\right)=\frac{\left(n_{j}-m_{ij}\right)\cdot S_{j}}{\sum_{k=1}^{J}\left(n_{k}-m_{ik}\right)\cdot S_{k}}.
\end{equation}
The likelihood $L$ for a complete sequence of $N$ discoveries $\{I(1),\,\ldots,\, I(N)\}$
can then be expressed as
\begin{equation}
L=\prod_{i=1}^{N}\frac{\left(n_{I(i)}-m_{iI(i)}\right)\cdot S_{I(i)}}{\sum_{j=1}^{J}\left(n_{j}-m_{ij}\right)\cdot S_{j}}.\label{eq:likelihood}
\end{equation}
The unknown parameters are the number of fields $n_{1},\,\ldots,\, n_{J}$,
whose likelihood can now be estimated based on the existing discoveries.
Using a brute force approach, the entire space of plausible values
for the variables $n_{1},\,\ldots,\, n_{J}$ has been sampled. The
values $n_{j}$ have been sampled between the number of existing fields
$m_{Nj}$ in the bin $j$ and up to a value $n_{j}^{upper}$, such
that the scenario with the largest likelihood according to equation
(\ref{eq:likelihood}) has $n_{j}^{max}=n_{j}^{upper}/2$ fields in
the bin $j$. Subsequently, the likelihood of each scenario (value
of the tuple $n_{1},\,\ldots,\, n_{J}$) has been normalized such
that the total likelihood of all generated scenarios equals one.

For the analysis of the discoveries of the North Sea oil fields, the number
of size bins has been fixed to $J=2$ splitting between dwarfs (1)
and giants (2) as described in section \ref{sub:Discoveries-modeled-as}.

The results shown in table \ref{likelihood-results} are coherent
with the intuitive expectation that discovering a new giant field
is unlikely and that future discoveries will mostly be made up of dwarf
fields. The likelihoods obtained for the carrying capacities of dwarfs
and giants have been used to constrain the logistic curve fitted to
the discoveries. Figure \ref{rate_of_discovery} shows a sample of
fitted logistic curves, each curve being weighted by the likelihood
of its carrying capacity given by equation (\ref{eq:likelihood}).

\subsubsection{Future production from discoveries\label{sub:Future-production-from}}

We now compute an expected oil production coming from future
discoveries, which requires to combine the steps described in sections
\ref{sub:Discoveries-modeled-as} and \ref{sub:Likelihood-function-for}.

The method described in section \ref{sub:Likelihood-function-for}
yields probabilities for the total number of fields (including the
not yet discovered fields) in each size bins (called a scenario).
However, this likelihood method does not give the time distribution
of future discoveries. We propose to use the likelihood function to
generate scenarios with their respective occurrence probability.

For a given scenario, the carrying capacity $K$ (= total number of
fields) is given for each size class. This is useful to resolve
the instability in fitting the logistic curve to the number of discovered
fields. The time distribution of the discoveries is then given for
each size class by the fitted logistic curve.

The actual size of a newly discovered field is generated according
to the size distribution of the existing fields in its size class.
The probability distribution function of field sizes in a given size
bin has been fitted by a stretched exponential function.

The production curve is computed based on the average production curve
of all existing fields in the same size category. The production curves
of the existing fields have all been normalized to a total production
of one and then have been averaged. This yields the typical production
profile including a one sigma confidence interval. For a new field,
this typical production curve is than multiplied by the size of the
field.

Superposing the production curves results in the expected production
curve from future oil fields for a given scenario.

As the total parameter space is too large to be sampled entirely,
a Monte Carlo technique is applied to compute the expected production
with confidence interval from future discoveries. In a nutshell, the
algorithm works as follows:
\begin{enumerate}
\item Draw a scenario (total number of fields in each size bin) based on its probability according to the likelihood function (\ref{eq:likelihood}). This is done by generating a random number $r$ between 0 and 1, and computing the scenario that is mapped to $r$ by the cumulative distribution function of all scenarios.
\item Compute the time distribution of new discoveries by fitting a logistic
curve for each size class.
\item For each discovery, generate a size and the resulting production curve
based on the size distribution and production curves of existing fields.
\item Superpose all the production curves.
\item Repeat and average over all drawn scenarios.
\end{enumerate}
The result is the expected production curve of future oil field discoveries.
The distribution of generated scenarios yields the confidence interval.

Last but not least, it has to be defined how the expected production from these future oil field discoveries is added to the extrapolated production from existing fields. The start of the simulation of new discoveries does not match up with the date of the latest production data, the reason being that the new fields (defined in section \ref{sub:Regular,-irregular-and}), which are already discovered but did not yet enter the decay phase, are not taken into account in the simulation as their final size is not known. The only meaningful way of treating the new fields is to consider them as a discovery. Therefore, the starting point in time of the simulated production resulting from new discoveries has been choosen as the date in the past where it matches the current production from
new fields. In order words, the extrapolated production from regular and irregular fields added to the production from simulated future discoveries (which is shifted into the past as to match the production from new fields) must be equal to the current (latest available data) total production from all fields.

\section{Results\label{sec:Results}}

\begin{figure}[H]
\includegraphics[width=1\columnwidth]{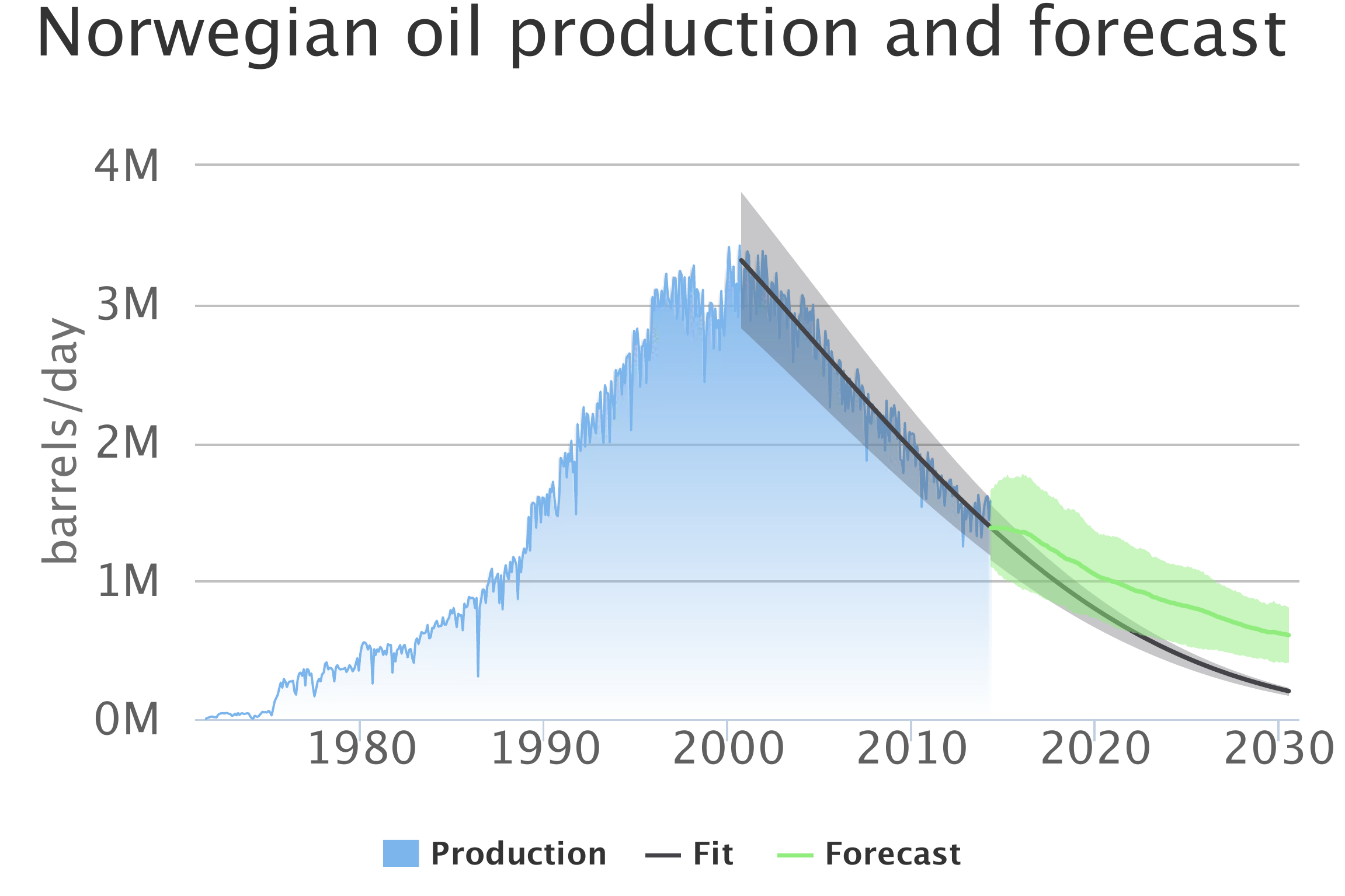}
\includegraphics[width=1\columnwidth]{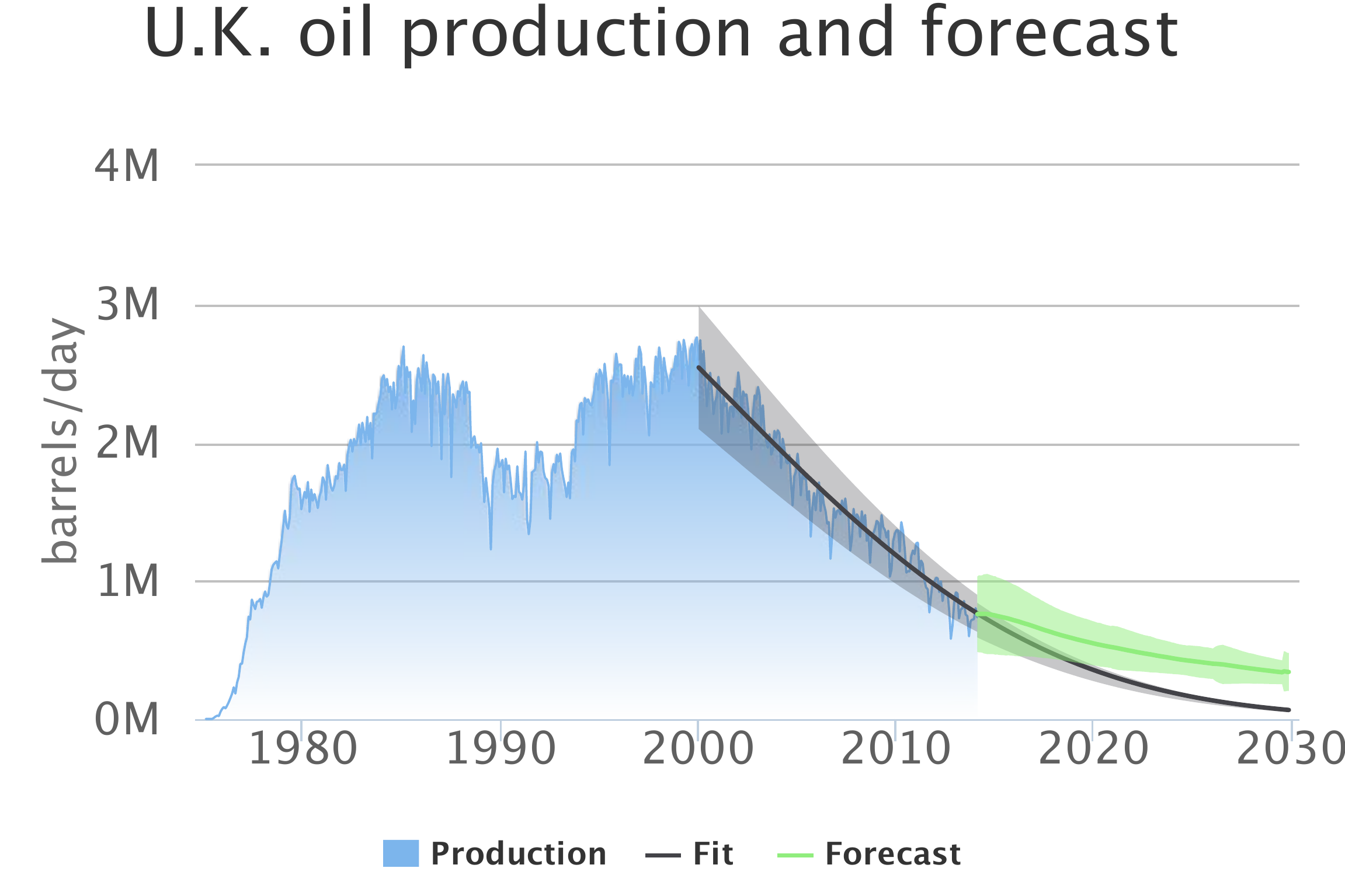} 
\caption{Monte-Carlo (green upper
continuous line with standard deviation bank starting in 2014 onward)
and fit forecast based on past production data (lower line and grey one standard deviation band) for Norway
(top) and the U.K. (bottom). In both cases, the Monte-Carlo model forecasts
a significantly slower decay than the fit by taking into account that
new fields will come in production.}
\label{no_uk_2013} 
\end{figure}

Based on the methodology described in section \ref{sub:Discovery-rate-of},
simulating future oil production was straightforward. For each country,
the existing oil field productions were extrapolated and the future discoveries were
simulated. Figure \ref{no_uk_2013} shows the average of 1000 simulations.
For each country, the non-symmetric shape of the production dynamics, which
contradicts the prediction based on Hubbert's standard approach,
is immediately noticeable.

The results in table \ref{forecast-14} show a striking difference
between the extrapolation of the fit and the Monte-Carlo model forecast. According to the
fit (extrapolation of aggregate production), Norway's future oil production
would decay much faster than in the Monte-Carlo case. The remaining
reserves estimated with the Monte-Carlo methodology are $45\%$ larger
then the estimate from the fit. This difference originates from two
different effects:
\begin{itemize}
\item The sum of the forecast of the individual existing fields is larger
than the extrapolation of the aggregate production.
\item The extrapolation of the aggregate production does not capture well
the discovery process of dwarf fields.
\end{itemize}
In the U.K., oil production faced a change of regime during the early
nineties due to technological innovation, giving rise to the inverted
\textquotedbl{}w shape\textquotedbl{} of the oil production profile.
However, this has not been an issue for us to extrapolate the decay of production
starting at the second peak. The difference between the Hubbert-based
methodology and the Monte-Carlo one is very similar to the Norwegian
case. The former underestimates the remaining oil reserves by about
$66\%$ compared to the latter.

Which of the two models is more trustworthy? Clearly, the implications
in adopting one methodology over the other are significant. The only
way to answer this question is to back-test them. In other words:
\textquotedbl{}What would each of the models have predicted, had they
been used in the past?\textquotedbl{} The next section addresses
this question and presents the validation step of our approach.

\section{Validation}

\begin{figure}[H]
\includegraphics[width=1\columnwidth]{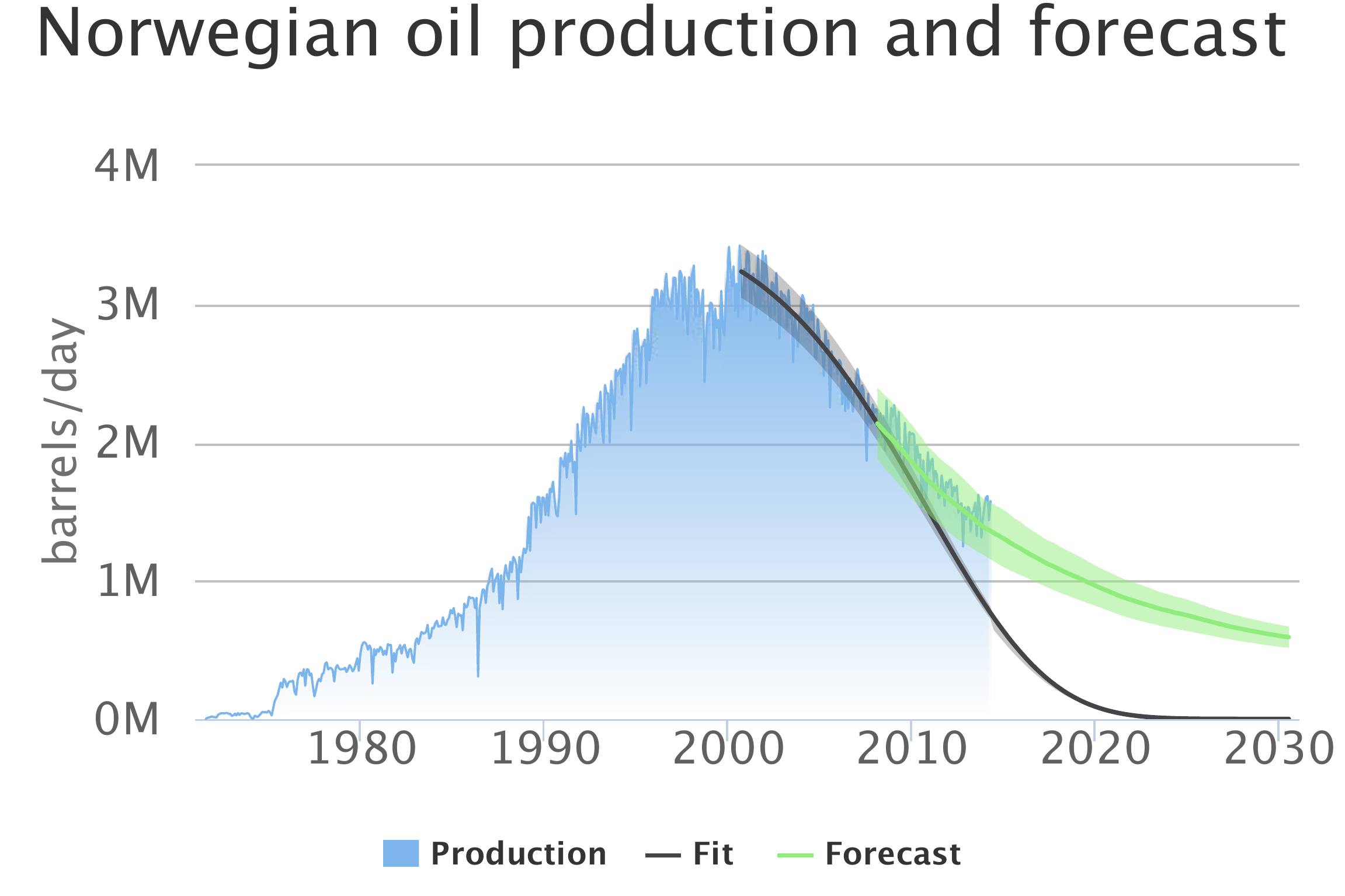}
\includegraphics[width=1\columnwidth]{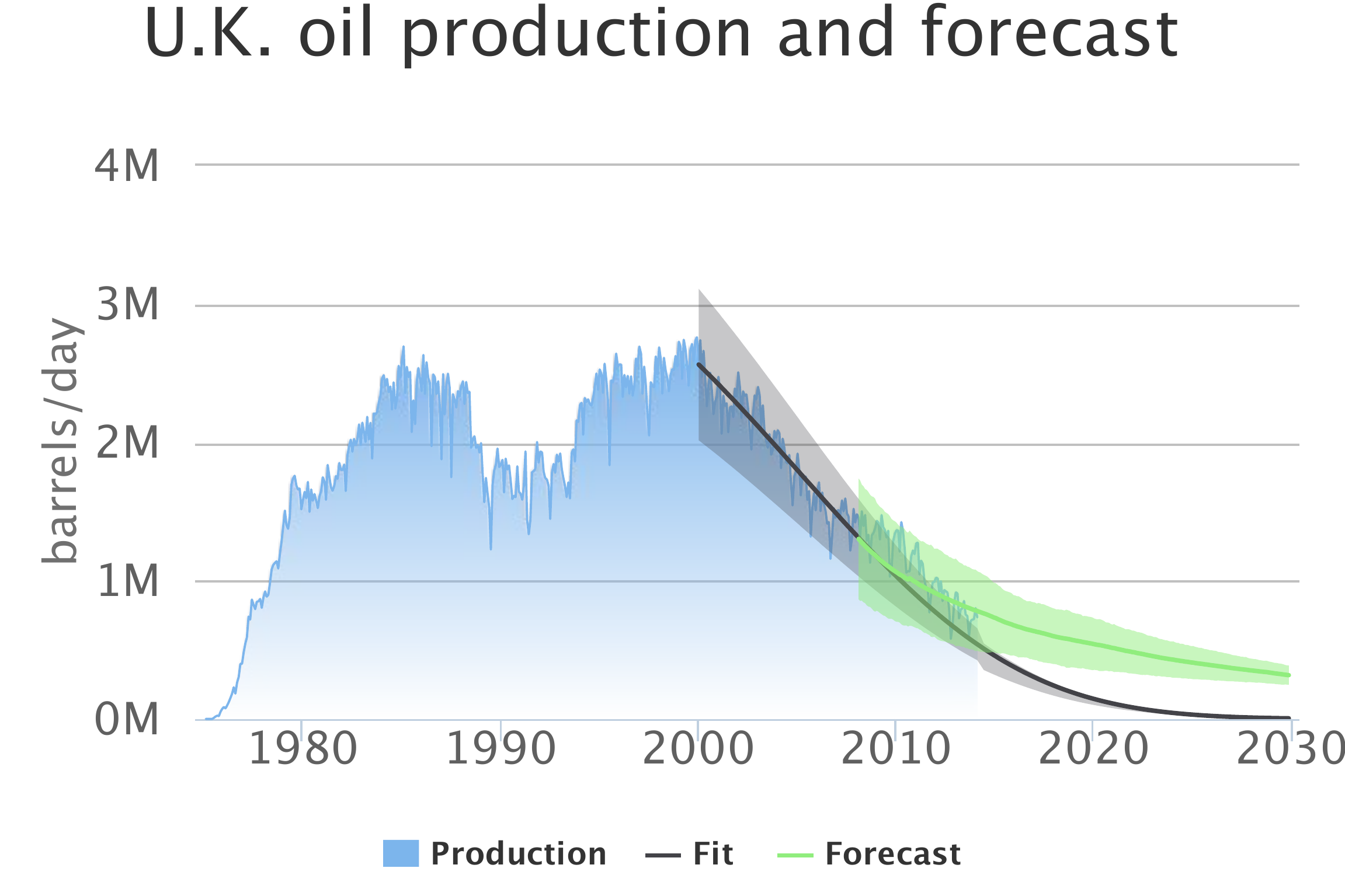} \caption{Monte-Carlo (green upper
continuous line with standard deviation bank starting in 2014 onward) and Hubbert forecast based 
on past production data up to 2008 (lower line and grey one standard deviation band)
for Norway (top) and the U.K. (bottom). The results can be
compared with the subsequent oil production (blue area). In both cases,
the Monte-Carlo methodology is more precise.}
\label{model-back-testing} 
\end{figure}

For both countries, namely Norway and the U.K., a back-test using
the data truncated in 2008 has been made. Before that date, too many
of the giant fields have not entered their decay phase for a sufficiently
long time to apply the extrapolation algorithm. Figure \ref{model-back-testing}
shows the results of these tests. Comparing the forecast of both models,
with the oil production of the subsequent 6 years, shows that the
predictive power of the Monte-Carlo model is significantly better than a simple
extrapolation of the aggregate past production. The Monte-Carlo model 
is found essentially right on target, while the extrapolation of past production
(``fit'') is under-estimating the realised production by 19\% and 16\% respectively
for Norway and the U.K. The following table \ref{backtest-08}
summarizes the difference between the two approaches for the back-testing
period.

While the error of the extrapolation of the past oil production method (``fit'') is not dramatic
over these six years from 2008 to 2014, the difference between it and the Monte-Carlo
approach becomes huge from 2014 to 2030, as shown in table \ref{forecast-08}.

The simple extrapolation decays too fast and entirely misses the fat
tails in the decay process of individual fields and the new discoveries.
Moreover, it must be noted that the simple extrapolation changed massively
between the back-test in table \ref{forecast-08} and the current
fit in table \ref{forecast-14} (520\% for Norway and 236\% for the
U.K.). In other words, the simple extrapolation is very unstable
in its forecast, while the Monte-Carlo forecasts remains very consistent
(less than 10\% change).

As can be seen in figure \ref{model-back-testing}, the actual production
of Norway during the back-testing period remained entirely within
the quite narrow $1\sigma$ interval of the Monte-Carlo methodology,
while totally breaking out of the $1\sigma$ interval of the simple
extrapolation. For the U.K. the Monte-Carlo methodology only performs
slightly better when considering the confidence interval, and the
confidence interval is much larger due to the uncertainty on future
discoveries and their production profile.

\section{Conclusion}

We have presented a Monte-Carlo based methodology to forecast future
oil production. By extending the oil production of current fields
into the future and modeling the discovery rate of new fields, the
future oil production of Norway and the U.K. could be forecasted.
These forecasts are significantly different from the ones obtained
with a standard extrapolation. Indeed, our model forecasts 45\% to
66\% more remaining oil reserves than the standard extrapolation.
The back-test performed on the time period between 2008 and 2014 confirmed
that the Monte-Carlo based model better captured the production dynamics.

The results suggest that it is highly likely that the decay
of Norwegian and U.K. oil production will be much slower then one
would expect from a standard extrapolation. Nonetheless, to maintain
current levels of oil consumption in the European Union, more of it
will have to be imported from outside Europe, as the imports from
Norway will vanish (currently accounting for 11\% of E.U. oil imports
\citepalias{Commission2014}) and the U.K. will need to import more
oil.

As shown in table \ref{needs}, at constant consumption, the Monte-Carlo
model predicts that in 2030 the E.U. with Norway will need to increase
its oil imports by $1.3$ million barrels. These imports will most
likely have to come from outside Europe, except for  
non-standard oil sources yet to be developed.

The present methodology can be applied to many other countries
and geological areas, as well as updated at the level of the global
oil production. Extensions to include the new wave of shale oil and
non-standard oil can in principle be considered and constitute
an interesting domain of application of our methodology for the future.

\vskip 0.5cm
\noindent
{\bf Acknowledgements}: This work derives from a Master thesis at ETH Zurich authored by 
Marc A. Del Degan (July 2012) under the supervision of the authors, 
entitled ``Analysis of peak oil with focus on Norwegian oil production''
(\url{http://www.er.ethz.ch/publications/MAS_Thesis_DelDegan_final_July12.pdf}), which
itself derived from the methodology of pricing social network companies such as Zynga developed
by three of the authors \cite{Forro2012}.

\bibliographystyle{model5-names}
\biboptions{authoryear}
\bibliography{bibliography.bib}

\newpage{}

\noindent 

\noindent 

\pagebreak

\noindent 

\noindent 

\pagebreak

\pagebreak

\noindent 

\pagebreak


\begin{table}[h]
\caption{Likelihood estimation for the number of dwarf (1) and giant (2) fields.
For Norway, our logistic fit suggests that up to two new giant fields could be
ultimately discovered. For the U.K., the prediction is more bleak, suggesting
that the most probable scenario is that one giant field will be found (which is most likely already discovered but classified as a new field).}
\centering %
\begin{tabular}{lllll}
 & $m_{N1}$ & $n_{1}\pm\sigma_{1}$ & $m_{N2}$ & $n_{2}\pm\sigma_{2}$\tabularnewline
\midrule Norway  & $24$ & $88.4\pm10.0$ & $52$ & $56.4\pm1.6$\tabularnewline
U.K. & $162$ & $208\pm11$ & $99$ & $100\pm0.4$\tabularnewline
\bottomrule  &  &  &  & \tabularnewline
\end{tabular}\label{likelihood-results}
\end{table}

\pagebreak


\begin{table}[H]
\caption{Remaining oil reserves until 2030 in barrels 
predicted by the extrapolation of the  \textbf{Fit}
of the past country production, and predicted by the Monte-Carlo \textbf{Model}.
The relative difference between these two predictions 
is defined by $\Delta=\frac{\textrm{Model}-\textrm{Fit}}{\textrm{Fit}}$.}
\centering %
\begin{tabular}{lllr}
 & Fit (barrels) & Model (barrels)  & $\Delta$ \tabularnewline
\midrule Norway & $130\cdot10^{6}$  & $188\cdot10^{6}$  & $45\%$\tabularnewline
U.K. & $59\cdot10^{6}$  & $98\cdot10^{6}$  & $66\%$ \tabularnewline
\bottomrule  &  &  & \tabularnewline
\end{tabular}\label{forecast-14}
\end{table}

\pagebreak


\begin{table}[H]
\caption{Extrapolation of past oil production (``fit'') and prediction
using the Monte-Carlo model are used on the data set truncated
in 2008. Their forecast for the period 2008-2014 is compared to
the actual realised production.}
\centering %
\begin{tabular}{llll}
 & Actual (barrels) & Fit (barrels) & Model (barrels)\tabularnewline
\midrule Norway & $133\cdot10^{6}$  & $108\cdot10^{6}$  & $130\cdot10^{6}$ \tabularnewline
U.K. & $79\cdot10^{6}$  & $66\cdot10^{6}$  & $75\cdot10^{9}$ \tabularnewline
\bottomrule  &  &  & \tabularnewline
\end{tabular}\label{backtest-08}
\end{table}

\pagebreak


\begin{table}[H]
\caption{Remaining oil reserves forecasted for the period 2014-2030 when using
the data truncated in 2008, according to the extrapolation of past oil production (``fit'') and 
the Monte-Carlo model. The relative difference between these two predictions 
is defined by $\Delta=\frac{\textrm{Model}-\textrm{Fit}}{\textrm{Fit}}$.}
\centering %
\begin{tabular}{lllc}
 & Fit (barrels) & Model (barrels) & $\Delta$ \tabularnewline
\midrule Norway 08 & $25\cdot10^{6}$  & $171\cdot10^{6}$  & $-584\%$\tabularnewline
U.K. 08 & $25\cdot10^{6}$  & $91\cdot10^{9}$  & $-264\%$ \tabularnewline
\bottomrule  &  &  & \tabularnewline
\end{tabular}\label{forecast-08}
\end{table}

\pagebreak


\begin{table}[H]
\caption{Oil import (bbl/day) at a constant consumption of 1.5M bbl/day for
the U.K. and 0.22M bbl/day for Norway. The import for the E.U. and
Norway is a lower bound based on the changes in the U.K and Norway. Negative
numbers for Norway represent exports.}
\centering %
\begin{tabular}{lllcc}
 & 2014 & 2020 & 2025 & 2030\tabularnewline
\midrule Norway & $-1.23\cdot10^{6}$  & $-0.88\cdot10^{6}$  & $-0.58\cdot10^{6}$  & $-0.43\cdot10^{6}$ \tabularnewline
U.K. & $0.7\cdot10^{6}$  & $0.9\cdot10^{6}$  & $1.0\cdot10^{6}$  & $1.1\cdot10^{6}$ \tabularnewline
E.U.+Norway & $9.8\cdot10^{6}$  & $10.45\cdot10^{6}$  & $10.85\cdot10^{6}$  & $11.1\cdot10^{6}$ \tabularnewline
\bottomrule  &  &  &  & \tabularnewline
\end{tabular}\label{needs}
\end{table}

\end{document}